\documentclass[%
 reprint,
nofootinbib,
 amsmath,amssymb,
 aps,
]{revtex4-2}

\allowdisplaybreaks
\usepackage{graphicx}
\usepackage{dcolumn}
\usepackage{bm}
\usepackage{textgreek}
\usepackage{afterpage}
\usepackage{subcaption}

\usepackage{multirow,graphicx}
\usepackage{makecell}

\usepackage[bottom]{footmisc}
\usepackage{xcolor}

\usepackage[colorlinks=true,linkcolor=blue,citecolor=blue,urlcolor=blue]{hyperref} 

\begin{document}

\def\apj{Astrophys.\ J.}
\def\apjl{Astrophys.\ J.\ Lett.}
\def\aap{Astron.\ Astrophys.}

\newcommand{\orcidicon}{\includegraphics[width=0.32cm]{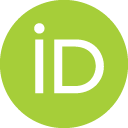}}
\newcommand{\orcid}[1]{\href{https://orcid.org/#1}{\orcidicon}}


\title{Ultralight Dark Matter Constraints from NanoHertz Gravitational Waves}

\author{Shreyas Tiruvaskar\orcid{0009-0003-9254-2199}$^1$ }
 \email{shreyas.tiruvaskar@pg.canterbury.ac.nz}

\author{Russell Boey\orcid{0009-0001-3893-8542}$^2$}%
\email{russell.boey@auckland.ac.nz}

\author{Richard Easther\orcid{0000-0002-7233-665X}$^2$}
\email{r.easther@auckland.ac.nz}
 
\author{Chris Gordon\orcid{0000-0003-4864-5150}$^1$}
\email{chris.gordon@canterbury.ac.nz}

\affiliation{\vspace{.3cm}  $^1$School of Physical and Chemical Sciences, University of Canterbury, New Zealand\\
  $^2$Department of Physics, University of Auckland,
Private Bag 92019, Auckland, New Zealand \vspace{.5cm}}

\date{\today}

\begin{abstract}
We investigate the impact of ultralight dark matter (ULDM) on the mergers of supermassive black holes (SMBH)  and the resulting stochastic gravitational wave background. ULDM is based on exceptionally light particles and yields galactic halos with dense central solitons.  This increases the drag experienced by binary SMBH, decreasing merger times and potentially suppressing gravitational radiation from the binary at low frequencies. We develop semi-analytic models for the decay of SMBH binaries in ULDM halos and use current pulsar timing array (PTA) measurements to constrain the ULDM particle mass and its fractional contribution to the dark matter content of the universe.  We find a median ULDM particle mass of \(7 \times 10^{-22}\) eV  and show that scaling relations suggest that the drag remains effective at relatively low ULDM fractions, which are consistent with all other constraints on the model. Consequently, future pulsar timing measurements will be a sensitive probe of any ULDM contribution to the overall dark matter content of the universe.  
\end{abstract}

\maketitle


\section{\label{sec:intro} Introduction }


Pulsar timing experiments provide direct evidence for the presence of a very low frequency gravitational wave background in the universe \cite{NANOGrav:2023gor,EPTA:2023fyk,Xu:2023wog,Miles:2024seg,Reardon:2023gzh}. These observations are broadly consistent with the signal expected from binary supermassive black hole  (SMBH) mergers that would follow the unions of their parent galaxies during the hierarchical growth of structure.

Gravitational wave emission will drive the final stage of the merger, but other dynamical processes must bring the SMBH into sufficient proximity for it to become effective. However, the obvious mechanisms which could contribute to this  appear to stall before gravitational wave emission becomes significant, creating the  ``final parsec problem''~\cite{Milosavljevic:2002ht}. Separately, the spectrum inferred from pulsar observations has less power at very low frequencies than expected for a stochastic background sourced by SMBH mergers. Both puzzles would be resolved if there is an additional source of drag at the centres of galaxies, which remains active as SMBH binaries reach the PTA band,  ensuring prompt mergers while removing energy which would otherwise source gravitational radiation. 

This paper explores Ultralight Dark Matter (ULDM) as a source of extra dynamical friction.  ULDM is a dark matter candidate which consists of extremely light scalar particles \cite{Preskill:1982cy,Abbott:1982af,Hu:2000ke,Lesgourgues:2002hk,Suarez:2013iw,Graham:2015ouw,Marsh:2015xka,Ferreira:2020fam}, motivated by the possible existence of very light states in string theory \cite{Svrcek:2006yi}. Astrophysically, it can resolve apparent discrepancies between predictions of minimal Cold Dark Matter (CDM)  and observations at subgalactic scales. Conversely, it matches the successful predictions of  CDM at large scales.  

A key feature of ULDM is the presence of solitonic cores at the centres of dark matter halos \cite{Schive:2014dra,Schive:2014hza}.  Their central density can be much higher than in a comparable CDM halo, significantly increasing the dynamical friction experienced by SMBH binaries.  Hui {\it et al.\/} \cite{Hui:2016ltb} extended Chandrasekhar's classic treatment of dynamical friction~\cite{Chandrasekhar:1943ys} to approximate the drag on a moving black hole in a uniform ULDM background. Inside a soliton,the situation is more complicated as the soliton responds coherently to the perturbation induced by the moving black hole. However, the dynamics of this system have been explored numerically for both single \cite{Lancaster:2019mde,Glennon:2023gfm,Wang:2021udl,Boey:2024dks} and binary  \cite{Annulli:2020lyc,Koo:2023gfm,Boey:2025qbo} black holes.

In particular, Ref.~\cite{Boey:2025qbo} showed that in large halos with  ULDM masses of $\sim 10^{-21}$eV the dynamical friction from the soliton would be significant at separations well below a parsec. In these simulations , the soliton is ``pinched'' by the SMBH, boosting the density and thus increasing the drag. In what follows, we develop a semi-analytic model of the dynamical friction on SMBH binaries at small separations. This includes the pinching and also allows for the possibility that ULDM does not provide all of the dark matter. This allows us to generate a library of gravitational strain spectra for different parameter combinations. Comparing this to the spectrum derived from the 15-year NANOGrav dataset, we can constrain the ULDM particle mass and fraction. 

This paper builds on previous studies of the impact of UDLM on the gravitational wave background from SMBH binaries. In particular, Aghaie {\em et al.\/} \cite{Aghaie:2023lan} impose constraints on the soliton and particle mass under the assumption that the dark matter consists entirely of ULDM and that the SMBHs are larger than the solitons. We generate a synthetic spectrum and  allow for mixed ULDM and CDM models, but their findings are consistent with ours. Tomaselli \cite{Tomaselli:2024ojz} conducted an analysis of pulsar timing residuals and similarly found a possible suppression at low frequencies due to ULDM drag, further motivating this work. Sarkar \cite{Sarkar:2025tiy} used dynamical friction from self-interacting ULDM to impose constraints on the interaction strength, providing a complement to this study. Separately, two of us (ST and CG) looked at the related issue of drag on SMBH binaries in self-interacting dark matter \cite{Tiruvaskar:2025lkq}, and this analysis builds on tools developed for that work.  


We find that for a particle of around $10^{-21}$eV the drag is sufficient to solve the final parsec problem and to reduce the long wavelength power in the stochastic gravitational wave background, such that the 68\%  credible interval of $\log_{10}(m/{10^{-22} \mathrm{eV}}) = 0.84^{+0.44}_{-0.48}$. Interestingly, the drag remains effective if the mass of the central soliton is reduced, as would happen with a mixture of  ULDM and regular CDM. As a result, our constraints are largely independent of the assumed ULDM fraction, provided a soliton forms. 

This apparent paradox arises because the stochastic gravitational wave background is primarily sourced by large black holes merging in the most massive halos.  In these systems, the mass of the SMBH binary is expected to be similar to the soliton mass, given the usual ULDM core-halo relationship~\cite{Schive:2014hza}.  This increases the impact of pinching with a fractional ULDM component, counteracting the lower soliton mass, which, together with changes to the set of halos which undergo mergers, largely offsets variations in the ULDM fraction. Consequently, while the full range of ULDM particle masses is potentially excluded in a ULDM-only universe \cite{Lazare:2024uvj}, much of the parameter range in which the drag improves the fit to PTA data is consistent with all other observations. 

That said, the improved fit to observations comes with several caveats. In particular, our drag model should be viewed as a well-informed hypothesis, rather than a rigorous model. In particular, SMBH mergers occur in the wake of the merger of their parent galaxies and their central solitons. The resulting soliton will be highly excited, but even the best ULDM-SMBH drag simulations begin with an unperturbed soliton \cite{Boey:2024dks}. Secondly, there have been no detailed studies of the soliton properties in fractional ULDM models, beyond Ref.~\cite{Schwabe:2020eac}, which uses a fixed single particle mass and halo mass but varies the mass fraction. Thirdly, the relevant scaling relations suggest that SMBH and soliton masses are similar in the systems which contribute to the stochastic background, but the core-halo relation does not account for the presence of SMBH. That said, these assumptions point to productive lines of inquiry for future studies of ULDM dynamics. 

Separately, the literature is rich with ``anomalies'' that resolved themselves as more data was obtained and analytic tools improved. Studies of the low-frequency stochastic gravitational wave background are in their infancy, and this analysis cannot ``discover'' ULDM with any more confidence than that with which the low-frequency deficit in the stochastic gravitational wave background has been detected. That said, if the low-frequency deficit is not confirmed by future analyses, the approach developed here could rule out a significant fraction of the currently permitted ULDM parameter space.


The structure of this paper is as follows. Section \ref{sec:uldm_background} introduces the Schr\"{o}dinger-Poisson equations, the semi-analytic model and the computation of the orbital decay rate from these profiles.  Section \ref{sec:strainspectrum}  describes the calculation of the gravitational strain spectrum, and  Section~\ref{sec:param_est} discusses the parameter estimation.  We discuss our results in Section \ref{sec:results} and conclude in Section~\ref{sec:discussion}.

 

\section{\label{sec:uldm_background} SMBH-ULDM Dynamics   } 


\subsection{Schr\"{o}dinger-Poisson System}

In the non-relativistic regime, ULDM is governed by the Schr\"{o}dinger-Poisson equations, or 
\begin{equation}
    i\hbar\dot{\psi} =-\frac{\hbar^2}{2m}\nabla^2\psi+m(\Phi_\mathrm{U}+V)\psi\, ,
    \label{SP1} 
\end{equation}
and 
\begin{equation}
    \nabla^2\Phi_\mathrm{U} =4\pi Gm|\psi|^2 \, .
    \label{SP2}
\end{equation}
Here $\psi$ is the ULDM field, $m$ is the ULDM particle mass, $\Phi_\mathrm{U}$ is the ULDM gravitational potential, and $V$ is that external gravitational potential sourced by the black hole binary. In mixed dark matter scenarios, we neglect the potential of the CDM halo as its scale radius is considerably larger than the soliton for most dark matter fractions we consider. As such, the potential is essentially constant and has no impact on the soliton profile.

ULDM solitons are ground state solutions to the Schr\"{o}dinger-Poisson equations.  Following Ref~\cite{Boey:2024dks} we consider only the impact of the soliton itself on drag, neglecting the impact of the outer halo. At separations relevant to the PTA frequency range, we would not expect any interactions between the binary and the halo outside of the soliton \cite{Bromley:2023yfi}. However, this does neglect the impact of interactions between the halo and the soliton, which induce a random walk on the soliton \cite{Schive:2019rrw,Chowdhury:2021zik} and may affect the binary dynamics.



\subsection{\label{subsec:dddt}  Orbital Decay Rate}

For a given black hole mass, we use the relationships found in Ref.~\cite{NANOGrav:2023hfp} (hereafter referred to as Agazie2023) to determine the corresponding stellar mass,
\begin{equation}
    \log_{10}\left(\frac{M_{\mathrm{BH}}}{ \mathrm{M}_\odot}\right) = \mu + \alpha_\mu \log_{10} \left( \frac{M_{\mathrm{bulge}}}{10^{11}\, \mathrm{M}_\odot} \right) + \mathcal{N}(0, \epsilon_\mu),
\label{mbh_mbulge}
\end{equation}
where \(\mu\), \(\alpha_\mu\), and \(\epsilon_\mu\) 
parameterise the relationship between the black hole mass ($M_\mathrm{BH}$) and the bulge mass ($M_{\mathrm{bulge}}$).
Here, we set \(\alpha_{\mu}\) to 1.1 following the fiducial value given in Table 2 of  Agazie2023  while $\mu$  varies in our analysis. The Gaussian random scatter is denoted  by \(\mathcal{N}(0, \epsilon_\mu)\); it has zero mean and its  standard deviation  \(\epsilon_{\mu}\) is also varied, as discussed in Section \ref{sec:param_est}. The fraction of the galaxy's stellar mass present in the stellar bulge is \(M_{\mathrm{bulge}}\), and it is related to the overall stellar mass of the galaxy \(M_\star\) by \cite{NANOGrav:2023hfp}
\begin{equation}
    M_{\mathrm{bulge}} = 0.615 \cdot M_{\star}.
\label{mbulge_mstar}
\end{equation}

The stellar mass is, in turn, used to find the halo mass, $M_h$, through the relationship  \cite{Girelli_2020},
\begin{equation}
    \frac{{M}_{\star}}{{M}_h}(z)=2A(z)\left[\left(\frac{M_h}{M_A(z)}\right)^{-\beta(z)}+\left(\frac{M_h}{M_A(z)}\right)^{\gamma(z)}\right]^{-1},
\end{equation}
where $z$ is the redshift, and we use the parameters from Table 3 of Ref.~{\cite{Girelli_2020}} and solve for $M_h$ numerically. Given a UDLM particle mass $m$, we find the soliton mass from the canonical core-halo mass relationship  \cite{Schive:2014hza},
\begin{equation}
\label{eq:has_a}
    M_c=\frac{1}{4 a^{1/2}}\left(\frac{\zeta(z)}{\zeta(0)}\right)^{1/6}\left(\frac{M_{\mathrm{ULDM}}}{M_{\mathrm{min},0}}\right)^{1/3}{M_{\mathrm{min},0}},
\end{equation}
\begin{equation}
\zeta(z)=\frac{18\pi^2+82(\Omega_\mathrm{M}(z)-1)-39(\Omega_\mathrm{M}(z)-1)^2}{\Omega_\mathrm{M}(z)}
\end{equation}
and 
\begin{equation}
M_{\mathrm{min},0}= \frac{32\pi\zeta(0)^{1/4}}{375^{1/4}(H_0m/\hbar)^{3/2}\Omega_\mathrm{M0}^{3/4}} \rho_{M0} \, ,
\end{equation}
where $a$ is the scale factor and $H_0$ is the present value of Hubble's constant.

We defined $M_{\mathrm{\mathrm{ULDM}}}=fM_h$ to accommodate models in which ULDM contributes only some of the dark matter. Naively, $f$ represents the ULDM fraction, or
\begin{equation}
\label{eq:fraction}
   f= \frac{\Omega_{\mathrm{ULDM}}}{\Omega_{\mathrm{ULDM}}+\Omega_{\mathrm{CDM}}}\ \times \frac{\Omega_\mathrm{DM}}{\Omega_\mathrm{M}}.
\end{equation} 
There is significant ambiguity in the core-halo relation itself  \cite{Schive:2014hza,Schwabe:2016rze,Mocz:2017wlg, Padilla:2020sjy,Burkert:2020laq,Nori:2020jzx,Mina:2020eik,Taruya:2022zmt,Zagorac:2022xic,Kendall:2023kit} and it has not been investigated for mixed   models. However, it is unlikely that the soliton mass will scale linearly with the ULDM fraction. On top of this, analyses of the core-halo relationship have neglected the baryonic component of the universe, although it is included in equation~(\ref{eq:fraction}). In practice, $f$ thus expresses the relationship between the expected and actual soliton masses and not the actual ULDM fraction. 

In simulations of halo collapse in mixed models \cite{Schwabe:2020eac} with $m=2.5\times 10^{-22}$eV and halo masses of ${\cal{O}}(10^9) M_\odot$ solitons do not form if $f \lesssim 0.3$. In the very largest halos (which are our primary interest), the central soliton will be particularly compact, and the lower bound on $f$ will presumably be smaller.  We take $f\ge 0.01$ in what follows, which is a somewhat arbitrary bound, but our results do not depend on its specific value. Note too that the definition of $f$ in Ref.~\cite{Schwabe:2020eac} does not have the factor of ${\Omega_\mathrm{DM}}/{\Omega_\mathrm{M}}$ as baryons are not included in that analysis.

A separate consideration in mixed dark matter models is whether a soliton is able to reasonably form through the dynamical relaxation within the age of the universe. Following the formalism in \cite{Bar:2021kti}, a rough criterion for this compares the radius in the initial halo that contains enough ULDM to form a soliton of the desired mass, $r_{supply}$, with the radius at which the relaxation timescale, given by \cite{Levkov:2018kau}
\begin{equation}
    \tau=\frac{0.7\sqrt{2}m^3\sigma^6}{12\pi^3G^2\rho^2\textrm{ln}(m\sigma R)},
\end{equation}
is less than the age of the galaxy, $r_{relax}$. Here $\sigma$ is the velocity dispersion, $\rho$ is the initial ULDM density, and R is the characteristic radius of the system. Under the condition $r_{relax}>r_{supply}$, a soliton is expected to form. 

This condition is not necessarily fulfilled for all of our parameter space - for black hole masses above $\sim 10^8 \textrm{M}_{\odot}$ and particle masses $\sim 10^{-22}$eV, a majority of the dark matter must be ULDM for these conditions to be fulfilled. However, this treatment is for halos that have grown via direct collapse rather than mergers, as we consider here. Moreover, as noted above $f$ need not be directly equivalent to the ULDM fraction so we continue to treat it as a free parameter. 

Unperturbed ULDM solitons have a universal profile, which is the ground state of the Schr\"{o}dinger-Poisson system and is approximated by \cite{Schive:2014dra}
\begin{equation}
    \rho(r)=\frac{1.9(m/10^{-23}\textrm{eV})^{-2}(r_c/\textrm{kpc})^{-4}}{(1+0.091(r/r_c)^2)^8}M_{\odot}\textrm{pc}^{-3} \, .
    \label{eq:Dens_profile}
\end{equation}
The core radius $r_c$ is the point at which the density is 50\% of the central density, and the mass contained within it is $M_c = 0.2408 M_s$, where $M_s$ is the total soliton mass. The usual scaling relations suggest that the ratio of the masses of the black hole and the soliton is largest in the most massive halos, which contribute most of the stochastic signal. However, since it is not known how the core-halo mass relation responds to the presence of the SMBH, we simply take the unmodified form.

We compute ${\mathrm dD}/{\mathrm dt}$ (where $D$ is the distance between the SMBH) by extending the approach of  Ref.~\cite{Boey:2025qbo}.  The expected drag is   \cite{Hui:2016ltb}, 
\begin{equation}
    F_{i}=\frac{4\pi G^2 M_{i}^2 \rho_{i}}{v_{i}^2} C_{i},
    \label{eq:Hui_friction}
\end{equation}
where subscript $i$ indicates values associated with one of the black holes, and $\rho_i$ is the density of the ULDM soliton at the position of the black hole. The coefficient of friction $C_{i}$ is  
\begin{equation}
    C_{i} = \textrm{Cin}({2k_{i}\tilde{r}_{i}})+\frac{\sin(2k_{i}\tilde{r}_{i})}{2k_{i}\tilde{r}_{i}}-1 + \cdots
    \label{Coefficient}
\end{equation}
where $\textrm{Cin}(x)=\int_0^x(1-\cos{t})/tdt$, $r_{i}=|\boldsymbol{r_{i}}|$, $k_{i}=m v_{i}/ \hbar$, $\tilde{r}_{i}$ is a cutoff  taken to be $\tilde{r}_{i} = \alpha r_{i}$ and $\alpha$ is a numerical  parameter less than unity \cite{Koo:2023gfm}. We fix $\alpha=0.4$, based on the simulations in Ref.~\cite{Boey:2025qbo}.

We assume the centre of mass of the binary and the soliton are both at the origin, giving $M_1\boldsymbol{r_1}+M_2\boldsymbol{r_2}=0$. The velocities of the black holes are thus 
\begin{equation}
    v_{i}=\sqrt{\frac{GM_{\textrm{enc}_{i}}}{r_{i}}+\frac{GM_{j}^3}{\left(M_{i} +{M_{j}}\right)^2r_{i}}},
    \label{eq:circmotion}
\end{equation}
where subscript $j$ indicates the other black hole and $M_{\textrm{enc}_{i}}$ indicates the mass of soliton enclosed within the black hole's position. This provides the torque,
\begin{widetext}
    \begin{align}
    \frac{\mathrm dL_{i}}{\mathrm dt}&=\frac{\mathrm dr_{i}}{\mathrm dt}\frac{\mathrm dL_{i}}{\mathrm dr_{i}}\nonumber\\&    
   = \frac{\mathrm dr_{i}}{\mathrm dt} \frac{M_{i}\sqrt{G}}{2} \times \left(\sqrt{\frac{M_{\textrm{enc}_{i}}+\frac{M_{j}^3}{\left(M_{i} + M_{j}\right)^2}}{r_{i}}}+\right.  \left. \frac{\mathrm dM_{\textrm{enc}_{i}}}{\mathrm dr_{i}}\sqrt{\frac{r_{i}}{M_{\textrm{enc}_{i}}+\frac{M_{j}^3}{\left(M_{i}+ {M_{j}}\right)^2}}}\right)\nonumber\\&=\frac{\mathrm dr_{i}}{\mathrm dt}\frac{M_{i}}{2}\left(v_{i}+\frac{4\pi r_{i}^2 \rho_{i} G}{v_{i}}\right),
    \label{eq:dLdtfull}
\end{align}
\end{widetext}
where in the final line we have made use of spherical symmetry to find ${\mathrm dM_{\textrm{enc}_{i}}}/{\mathrm dr_{i}}$. 
This can be equated to the torque due to dynamical friction, $-F_{i}r_{i}$, to find  ${\mathrm dr_{i}}/{\mathrm dt}$,
\begin{equation}
    \frac{\mathrm dr_{i}}{\mathrm dt}=-\frac{8\pi G^2 M_{i} \rho_{i}}{v_{i}^3+4\pi r_{i}^2 \rho_{i} Gv_{i}} Cr_{i} \, .
    \label{eq:rdotfull} 
\end{equation}
Summing these two terms gives ${\mathrm dD}/{\mathrm dt}$.

The density in equation (\ref{eq:rdotfull}) depends on the soliton profile, which is ``pinched'' by the pair of SMBH. We estimate the modified profile by assuming that the total mass of the binary is located at the soliton centre.  We  additionally assume time-independence (in the radial direction) with the ansatz  $\psi(\boldsymbol{r},t)=e^{i\beta t}F(r)$ and $\Phi_{\textrm{U}}(\boldsymbol{r},t) = \phi(r)$ and the resulting  Schr\"{o}dinger-Poisson equation is
\begin{align}
    F''(r)+\frac{2}{r}F'(r)-2\tilde{\phi}(r)F(r)+2\frac{M_{BH}}{r}F(r)=0,
    \label{Dimensionless_prof}\\
    \tilde{\phi}''(r)-4\pi F^2(r)+\frac{2}{r}\tilde{\phi}'(r)=0,
    \label{Dimensionless_pot}
\end{align}
where $\tilde{\phi}(r)=\phi(r)+\beta$ and $F'(r)={\mathrm dF}/{\mathrm dr}$. The equation is solved via the shooting method, with $F(0)=1$, $F'(0)=\tilde{\phi'}(0)=0$, and $F(r_{\textrm{max}})={\phi}(r_\textrm{max})=0$, where $r_{\textrm{max}}$ is a cut-off radius where the soliton profile is negligible. 
The mass of the soliton can then be found by integrating the profile, and this in turn gives the mass ratio between the black hole and the soliton. 

If $e^{i\beta t}F(r)$ solves the radial Schr\"{o}dinger-Poisson equations,  the scaled profile $e^{i\gamma \beta t}\gamma F(\sqrt{\gamma}r)$ is also a solution. After restoring dimensionality, $\gamma$ is fixed by requiring the integrated mass (recalling that the density is proportional to $|\psi|^2$) matches the desired soliton mass, and the black hole to soliton mass ratio is invariant with $\gamma$.  By varying the black hole mass, we can solve iteratively for the profile\footnote{In practice, we generated profiles for mass ratios between 0 and 82.15 and interpolated between them.} and, as noted earlier, the pinching is significant when the black hole mass approaches or exceeds the soliton mass.

\begin{figure}[tb!]
    \centering
    \includegraphics[width=\linewidth]{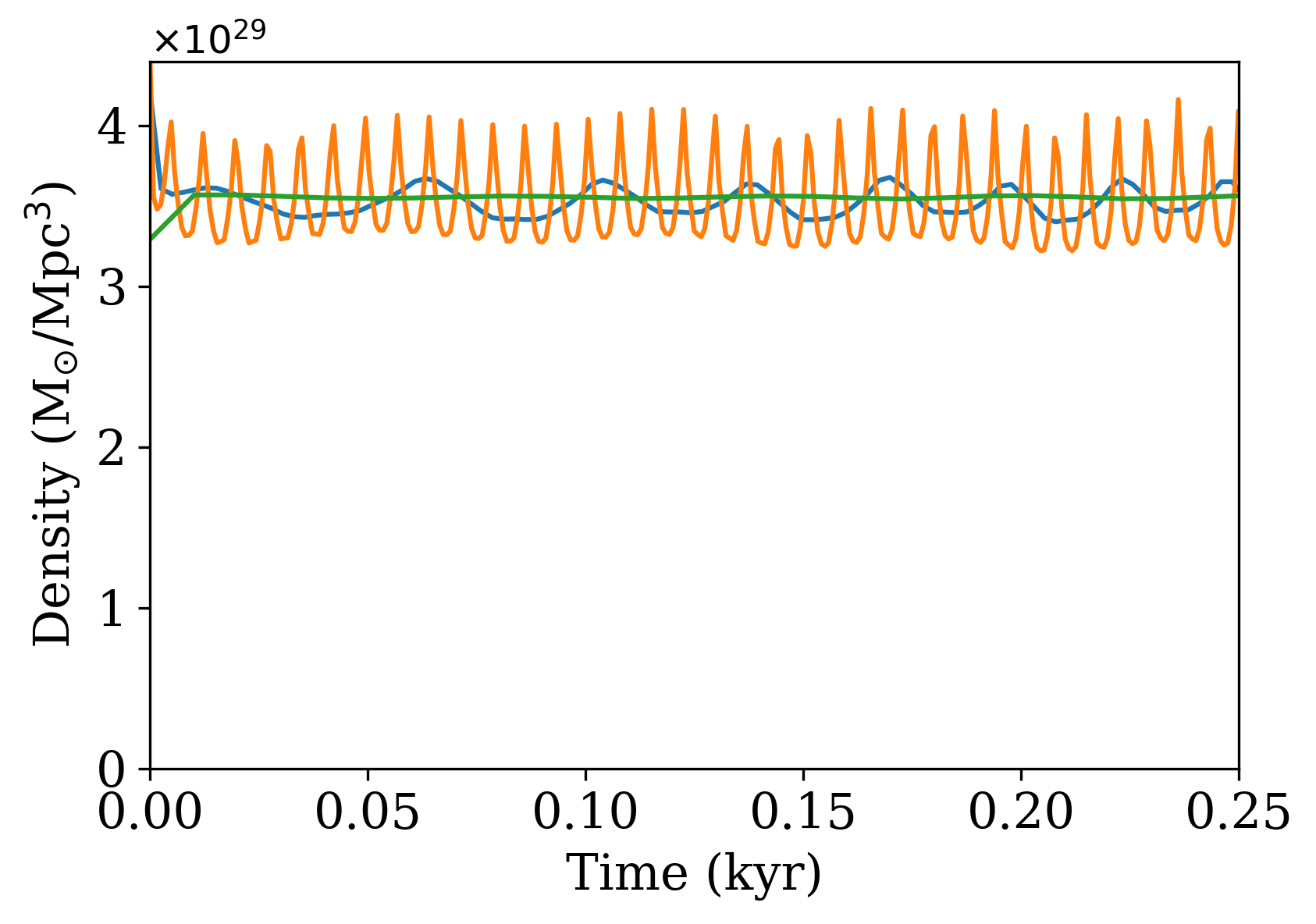}
     \includegraphics[width=\linewidth]{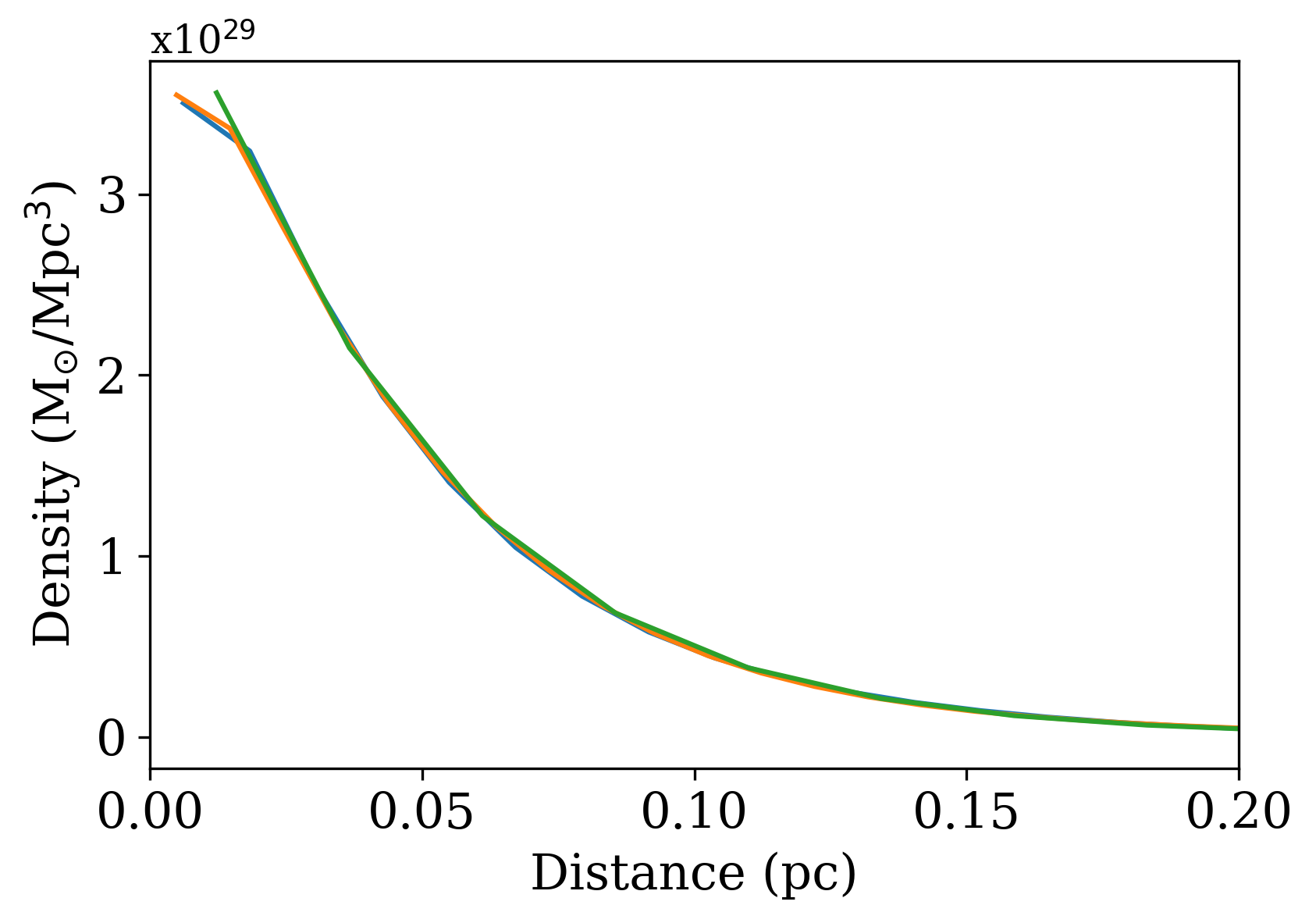}
    \caption{(Top) Central densities as a function of time for an equal mass binary (blue), a binary with mass ratio  $1:3$ (orange), and a stationary central black hole (green); the total SMBH mass is $10 M_s$. For the equal mass case, the initial radius is 0.015pc; the unequal mass case has 0.03pc and 0.01pc; $r_c\approx  0.0296$pc. The central densities of the binaries oscillate but are near the single black hole value. (Bottom) Radial density profiles at   $ t\sim 0.264$kyr.}
    \label{fig:single_v_binary}
\end{figure}

We tested these profiles by simulating binaries in circular orbits inside soliton profiles with {\sc AxioNyx} \cite{Schwabe:2020eac}, using the approach of Ref.~\cite{Boey:2025qbo}. There is good overlap for tight binaries, but, unsurprisingly, if the binary is outside the core radius of the pinched soliton, this approach breaks down. Figure \ref{fig:single_v_binary} shows the central density and representative radial profiles for examples when the total SMBH mass is $10M_s$. 
Given the pinched density profile, we can compute ${\mathrm dD}/{\mathrm dt}$ at any separation. 

This model ignores the backreaction on the soliton structure that can be induced by the binary, which can cause oscillations in the orbital decay of the binary. However,  with equal mass black holes the average decay is well-described by the semi-analytic formula \cite{Boey:2025qbo}. That said, when one black hole is much larger than the other and both black hole masses are small compared to the soliton the excitation of dipole terms in the soliton can significantly slow the inspiral \cite{Zhang:2026mse}, but the systems that dominate our signal do not satisfy these conditions.



\subsection{\label{subsec:simp} Simple Model}

To assess the sensitivity of the results to our model for the dynamical friction, we also consider a simplified model, without pinching, developed by Koo {\em et al.\/} \cite{Koo:2023gfm} and extended in Ref.~\cite{Boey:2025qbo}. The unpinched soliton profile is given by equation (\ref{eq:Dens_profile}). In this case, the SMBH separations are much smaller than $r_c$ when gravitational wave emission occurs, so the black hole velocities are
\begin{equation}
    v_{i}=\sqrt{\frac{GM_{j}^3}{\left(M_{i} +{M_{j}}\right)^2r_{i}}}.
    \label{eq:circmotion_simp}
\end{equation}
since we can neglect the ULDM enclosed by their orbit.

The expression for $C_{i}$ also simplifies when $k_{i}\tilde{r}_{i}\ll1$, and a Taylor expansion provides $C_i=(k_{i}\tilde{r}_{i})^2/3$. Finally, the ULDM density near the black holes is close to the central density, so we set $\rho_1=\rho_2=\rho(0)$. For this case 
\begin{equation}
    \frac{\mathrm dr_{i}}{\mathrm dt}=\frac{8\pi G^{3/2}M_{i}(1+M_{i}/M_{j})\rho(0)\alpha^2m^2r_{i}^{3.5}}{3\hbar^2\sqrt{M_{j}}},
    \label{eq:drdt_simp}
\end{equation}
and we again find ${\mathrm dD}/{\mathrm dt}$ by summing  these two contributions.

\subsection{Scaling expectations \label{subsec:scaling}}

In both models ${\mathrm dD}/{\mathrm dt}$ depends much more strongly on the ULDM particle mass $m$ than on $f$. Via the standard scaling relations \cite{Schive:2014hza}  the simple model has $r_c\propto m^{-1}f^{-1/3}$, $\rho(0)\propto m^{-2}r_c^{-4}$, and ${\mathrm dD}/{\mathrm dt}\propto \rho(0)m^2$. As such, 
\begin{equation}
    \frac{\mathrm dD}{\mathrm dt}\propto m^4f^{4/3}. 
    \label{eq:scaling_simplified}
\end{equation}
The scaling relationships in the realistic model are not as clean, and in that case, the effect is counter-intuitively enhanced by the soliton mass decreasing as $m$ increases, since $M_c\propto m^{-1}$. This increases the pinching for a fixed black holes mass (and thus fixed halo mass, given our other assumptions), boosting the central density and ${\mathrm dD}/{\mathrm dt}$ relative to the simple model. Conversely, increasing $f$ increases the soliton mass, which decreases the pinching (since $M_c \propto f^{1/3}$), so the density scales more slowly than in the simple model.  The outcome is that   ${\mathrm dD}/{\mathrm dt}$ and the decay timescale are much more sensitive to $m$ than to $f$ for both models. This will be reflected in our results, which allow a wide range of $f$ but pick out a narrower range of $m$, particularly with the realistic model. 

\section{Gravitational strain spectrum \label{sec:strainspectrum}}

The gravitational strain of a single binary source is given in equation~(6) of  Agazie2023 as
\begin{equation}
    h_\mathrm{s}^2(f_{\mathrm{gw}})  = \frac{32}{5c^8} \frac{(G\mathcal{M})^{10/3}}{ d_c^2} (2\pi f_{\mathrm{orb}})^{4/3},
\end{equation}
where \(d_c\) is the comoving distance to the source, which is a function of redshift \(z\). The binary chirp mass is denoted by \(\mathcal{M}\), \(f_{\mathrm{gw}}\)  
is the observer-frame gravitational wave frequency, and \(f_{\mathrm{orb}}\) is the source-frame orbital frequency of the binary. The source frame gravitational wave frequency is double the binary orbital frequency, \(f_{\mathrm{gw-source}} = 2 f_{\mathrm{orb}}\). If the source is at redshift \(z\), these frequencies are related by
\begin{equation}  
        f_{\mathrm{gw}} = \frac{f_{\mathrm{gw-source}}}{(1+z)}  
        = \frac{2f_{\mathrm{orb}}}{(1+z)}.
    \label{eq: freq source obs}
\end{equation}
The chirp mass of the binary is given by equation~(1) of Agazie2023,
\begin{equation}
    \mathcal{M} = \frac{M q^{3/5}}{(1+q)^{6/5}},
\end{equation}
where \(q\) is the mass ratio defined as \(M_2/M_1\) with \(M_1\) being the primary (heavier) black hole mass and \(M_2\) being the secondary (lighter) black hole mass, and the total mass of the binary is (\(M=M_1 + M_2\)).

The total gravitational strain can be obtained by adding the individual binary strains using  equation~(5) of Agazie2023, 
\begin{equation}
    h_c^2(f_{\mathrm{gw}}) = \int dM \, dq \, dz \frac{\partial^4 N}{\partial M \partial q \partial z \partial \ln f_{\mathrm{orb}}} h_s^2(f_{\mathrm{orb}}),
\label{hc}
\end{equation}
where \(h_c(f_{\mathrm{gw}})\) is the total characteristic strain of the GWB over a logarithmic frequency interval. Following equation~(7) of  Agazie2023, we  write
\begin{equation}
    \frac{\partial^4 N}{\partial M \partial q \partial z \partial \ln f_{\mathrm{orb}}} = \frac{\partial^3 \eta}{\partial M \partial q \partial z} \cdot \tau(f_{\mathrm{orb}}) \cdot 4 \pi c \, (1 + z) \, d_c^2
\label{number}
\end{equation}
where $c$ is the speed of light. The differential volumetric number density of binaries is \(\partial^3 \eta/\partial M \partial q \partial z\) and depends on \((M, q, z)\). It gives the number of binaries of mass \(M\), mass ratio \(q\), and redshift \(z\) that would be present in a unit volume and is calculated below. 

The binary hardening timescale is the time (measured in the source frame, i.e. the centre of mass rest-frame of the SMBH binary) spent in a given logarithmic frequency interval is denoted \(\tau(f_{\mathrm{orb}})\). Following equations~(44-48) from \cite{Tiruvaskar:2025lkq}, \(\tau(f_{\mathrm{orb}})\) can be written as
\begin{equation}
    \tau(f_{\mathrm{orb}}) \equiv \frac{f_{\mathrm{orb}}}{\dot{f}_{orb}}  = -\frac{3}{2}\frac{D}{\dot{D}}\, .
\end{equation}
The total characteristic strain is then 
\begin{equation}
    \begin{aligned}
        h_c^2(f_{\mathrm{gw}}) = \int dM &\, dq \, dz  \frac{\partial^3 \eta}{\partial M \partial q \partial z} \cdot \left(-\frac{3}{2}\frac{D}{\dot{D}}\right) \\
        &\times  4 \pi c \, (1 + z) \, d_c^2 \\
        &\times \frac{32}{5c^8} \frac{(G\mathcal{M})^{10/3}}{d_c^2} (2\pi f_{\mathrm{orb}})^{4/3}.
    \end{aligned}
\label{strain_integral}
\end{equation}
For the full model \(\dot{D}\) is  given by equation~\eqref{eq:rdotfull} and by equation~\eqref{eq:drdt_simp}  for the simple model -- given the  \({\mathrm dr_{i}}/{\mathrm dt}\)  we add them to get \(\dot{D}\). 

Binary separation \(D\) is related to the observed gravitational wave frequency \(f_{\mathrm{gw}}\), total binary mass \(M\), and redshift \(z\) through equations~\eqref{eq:kepler} and \eqref{eq: freq source obs}. For each combination of (\(M, q, z\))  we evolve our binary system in 300 steps using \(\dot{D}\), starting from a separation of \(D = 1 \mathrm{pc}\), down to \(D = \mathrm{ISCO}\), where the ISCO is the innermost stable circular orbit, given by \({6GM}/{c^2}\). At each step of the evolution we have a specific value of \(D\) which gives us the corresponding \(f_{\mathrm{orb}}\), \(f_{\mathrm{gw}}\), and the source strain \(h_s^2(f_{\mathrm{orb}})\). This allows us to numerically evaluate the integral in equation~\eqref{hc} for each PTA frequency bin. Following Agazie2023, we set the integration limits to \(M: (10^4 \mathrm{M}_\odot,10^{12}  \mathrm{M}_\odot)\), \(q: (0.001, 1)\), and \(z: (0.001, 10)\). We also require that contributions to the total strain are from binaries that merge within the age of the universe, ensuring that we only consider systems for which fthe inal parsec problem is actually solved in our model. 
 
We follow equation~(22) from  Agazie2023 to calculate the differential volumetric number density of the binaries,
\begin{equation}
    \frac{\partial^3 \eta}{\partial M\, \partial q\, \partial z} = \frac{\partial^3\eta_{\text{gal-gal}}}{\partial M_{\star 1}\, \partial q_{\star}\, \partial z} \frac{\partial M_{\star 1}}{\partial M} \frac{\partial q_{\star}}{\partial q},
\end{equation}
where \(M_{\star 1}\) is the stellar mass of the primary merging galaxy and \(q_{\star}\) is the stellar mass ratio of the secondary to the primary galaxy. The relation between \(M_{\star 1}\) and \(M_1\) can be evaluated from equations~\eqref{mbh_mbulge} and \eqref{mbulge_mstar}, with  we set \(\alpha_\mu\) to 1.1. Using these relations \(\partial M_{\star 1}/\partial M\) and \(\partial q_{\star 1}/\partial q\) are \cite{Tiruvaskar:2025lkq}
\begin{equation}
    \begin{aligned}
        \frac{\partial M_{\star 1}}{\partial M} &= \frac{1}{\alpha_\mu} \frac{M_{\star 1}}{M} \\
        \frac{\partial q_{\star}}{\partial q} &= \frac{1}{\alpha_\mu} q_{\star}^{1 -\alpha_\mu}.
    \end{aligned}
\end{equation}
and \(M_{\star 1}\) and \(q_{\star}\) can be written in terms of \(M_1\) and \(M_2\), which in turn can be expressed in terms of \(M\) and \(q\).

For \(\partial^3\eta_{\text{gal-gal}}/\partial M_{\star 1}\, \partial q_{\star}\, \partial z\), we refer to equation~(13) from  Agazie2023,  
\begin{equation}
    \frac{\partial^3 \eta_{\text{gal-gal}}}{\partial M_{\star 1} \, \partial q_{\star} \, \partial z} =
    \frac{\Psi(M_{\star 1}, z')}{M_{\star 1} \ln(10)} \cdot
    \frac{P(M_{\star 1}, q_{\star}, z')}{T_{\text{gal-gal}}(M_{\star 1}, q_{\star}, z')}
    \cdot \frac{\partial t}{\partial z'}.
\label{d3eta}
\end{equation}
Here, \(z'\) is the redshift at the time before the galaxy merger starts. Everything in the RHS is calculated at this pre-merger redshift \(z'\). The redshift in the LHS is the redshift after the merger is complete, i.e., \(z' = z'[t]\) and \(z = z[t + T_{\text{gal-gal}}]\). 

To calculate \(\Psi(M_{\star 1}, z')\), \(P(M_{\star 1}, q_{\star}, z')\), and \(T_{\text{gal-gal}}(M_{\star 1}, q_{\star}, z')\), we use equation~(15), equation~(16), equation~(19), and equation~(20) from Agazie2023:
\begin{equation}
    \Psi(M_{\star 1}, z') = \ln(10) \, \Psi_0 \left[ \frac{M_{\star 1}}{M_{\psi}} \right]^{\alpha_{\psi}} \exp\left( -\frac{M_{\star 1}}{M_{\psi}} \right)
\end{equation}
where
\begin{equation}
    \begin{aligned}
        &\log_{10} \left( \Psi_0 / \mathrm{Mpc}^{-3} \right) = \psi_0 + \psi_z \cdot z', \\
        &\log_{10} \left( M_{\psi} / M_{\odot} \right) = m_{\psi,0} + m_{\psi,z} \cdot z', \\
        &\alpha_{\psi} = 1 + \alpha_{\psi,0} + \alpha_{\psi,z} \cdot z'.
    \end{aligned}
\label{gsmf}
\end{equation}
The values and prior ranges of all these parameters are given in Table \ref{table_param_values}. 

The galaxy pair fraction \(P(M_{\star 1}, q_{\star}, z')\) and the galaxy merger time \(T_{\mathrm{gal-gal}}(M_{*1}, \, q_{*}, \, z')\) are calculated using equations (19) and (20) of Agazie2023. We follow their analysis and use the same values for the parameters of the galaxy pair fraction and merger time, which can be found in Table B1 of Agazie2023. 
%

\begin{table}[htbp]
    \centering
    {
    \renewcommand{\arraystretch}{1.5}
    \begin{tabular}{|c|c|c|}
        \hline
        \textbf{Symbol} & \textbf{Fiducial Value} & \textbf{Priors} \\
        \hline
        $\mathrm{log_{10}}(m_{22})$        & $-$     & $\mathcal{U}(-2.5, 2.5)$ \\
        $\mathrm{log_{10}}(f)$   & $-$  & $\mathcal{U}(-2, 0)$ \\
        \hline
        $\psi_0$        & $-$     & $\mathcal{N}(-2.56, 0.4)$ \\
        $\psi_z$         & $-0.60$      & $-$ \\
        $m_{\psi,0}$    & $-$     & $\mathcal{N}(10.9, 0.4)$ \\
        $m_{\psi,z}$    & $+0.11$      & $-$ \\
        $\alpha_{\psi,0}$ & $-1.21$     & $-$ \\
        $\alpha_{\psi,z}$ & $-0.03$     & $-$ \\
        \hline
        $\mu$           & $-$     & $\mathcal{N}(8.6, 0.2)$ \\
        $\alpha_\mu$    & $+1.10$      & $-$ \\
        $\epsilon_\mu$  & $-$     & $\mathcal{N}(0.32, 0.15)$ dex \\
        $f_{\star, \mathrm{bulge}}$ & $+0.615$ & $-$ \\
        \hline
    \end{tabular}
    }
\caption{Priors and fiducial values for the model parameters}
\label{table_param_values}
\end{table}

\begin{figure*}[tbp]
  \centering
  \vspace{1cm}
  \includegraphics[width=\textwidth]{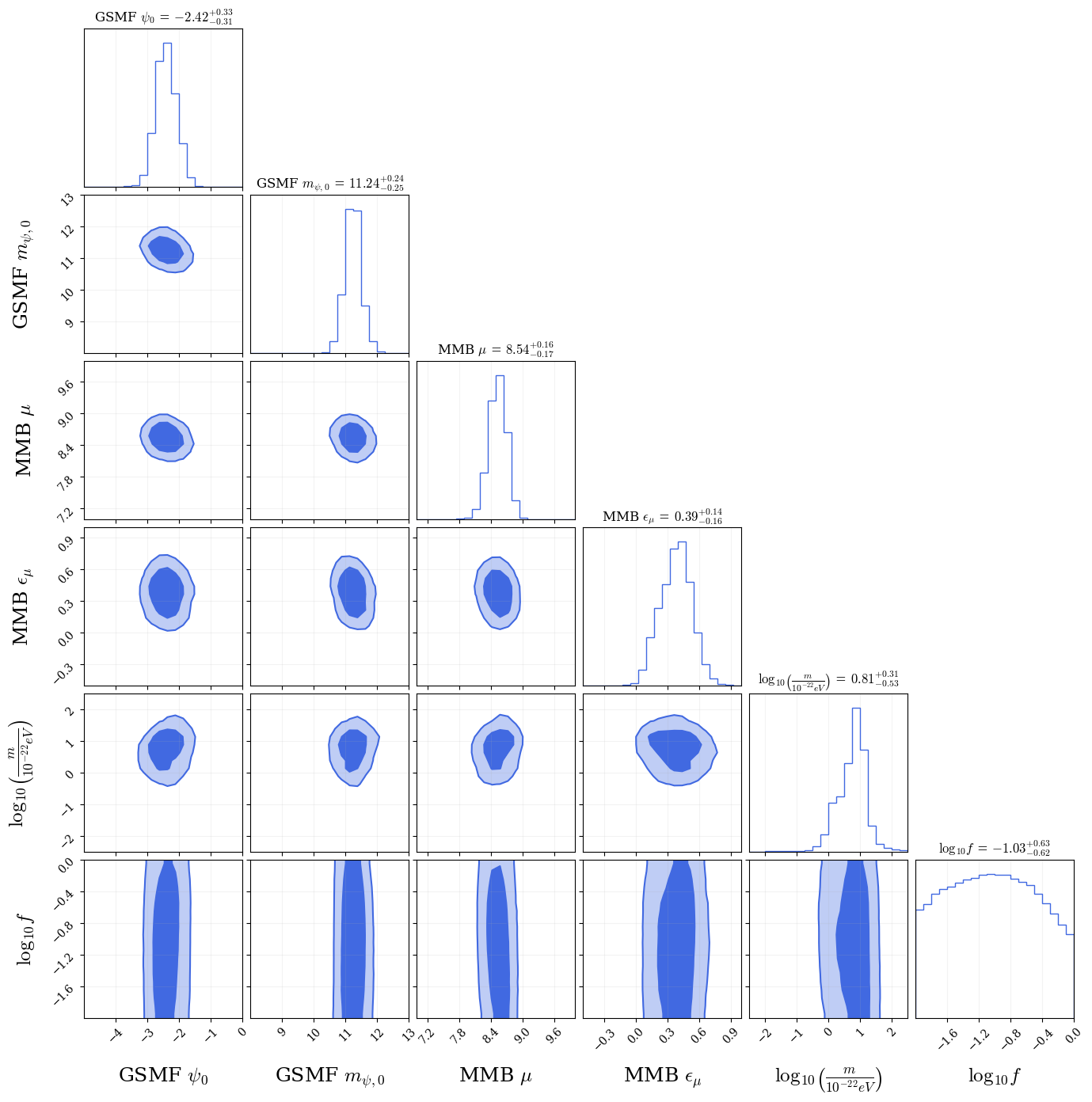}
  \caption{ Posterior distribution of model parameters with 68\% and 95\% confidence interval contours for the realistic model. Reported values correspond to the medians and their 68\% credible intervals.   }
  \vspace{1cm}
  \label{fig:posterior1}
\end{figure*}

\begin{figure*}[tbp]
  \centering
   \vspace{1.2cm}
  \includegraphics[width=\textwidth]{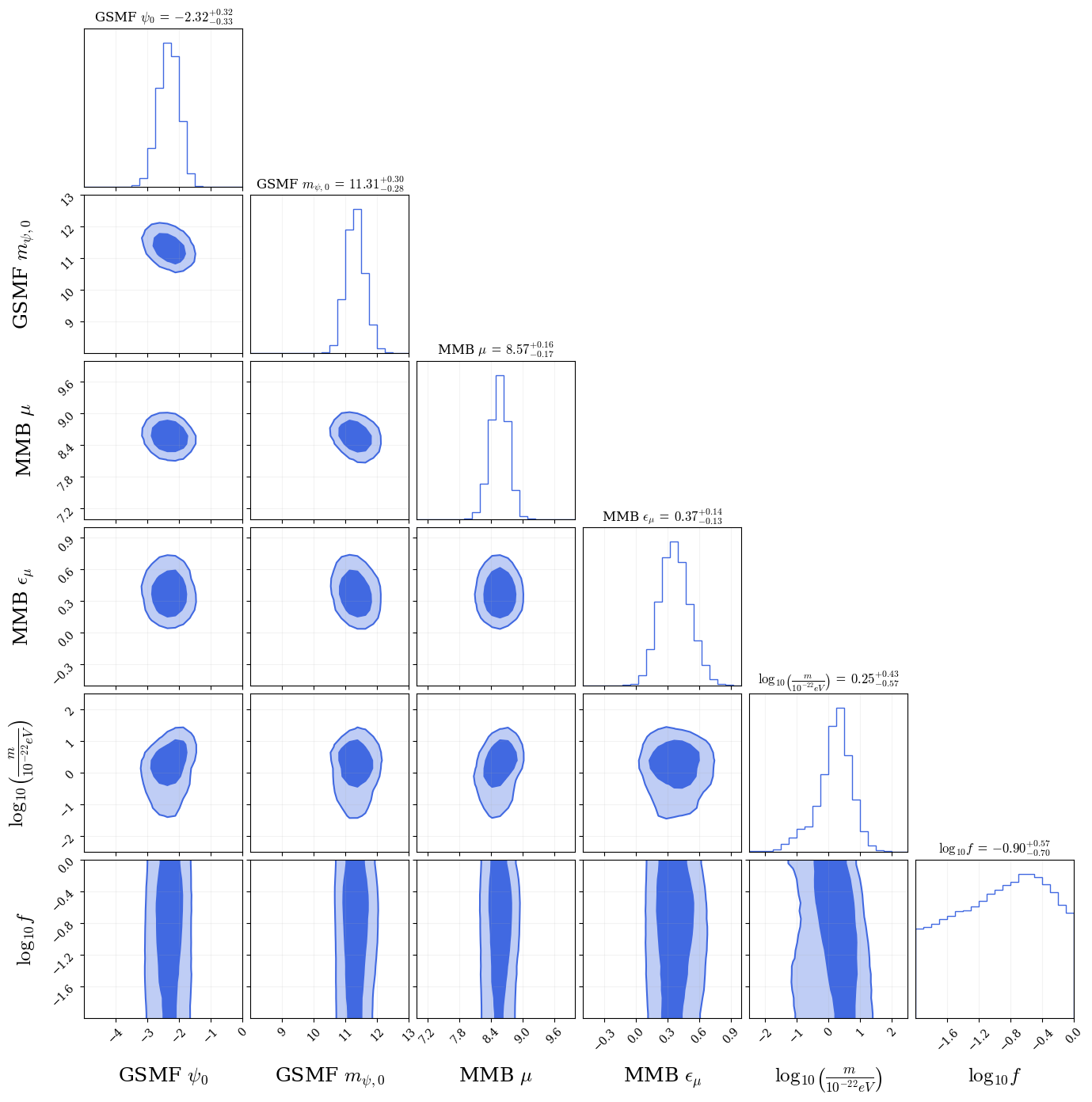}
  \caption{Posterior distribution for the simple model; other settings match those of Figure~\ref{fig:posterior1}. }
   \vspace{1.2cm}
  \label{fig:posterior2}
\end{figure*}

The last term in the RHS of equation~\eqref{d3eta} is \(\partial t/\partial z'\),  can be calculated using the redshift-time  relation
\begin{equation}
    \frac{\mathrm{d}t}{\mathrm{d}z'} = \frac{1}{(1 + z') H(z')}.
\end{equation}
where
\begin{equation}
    H(z) = H_0 \sqrt{\Omega_m (1+z)^3 + \Omega_{\Lambda}}.
\label{Hz}
\end{equation}
and  \(\Omega_{\Lambda}\) is the dark energy fraction. Following NANOGrav (see Agazie2023) we use the WMAP9 values for \(H_0\), \(\Omega_m\), and \(\Omega_{\Lambda}\),  \(H_0=69.33\,\mathrm{km\, s^{-1}\, Mpc^{-1}}\), \(\Omega_m = 0.288\) and \(\Omega_{\Lambda} = 0.712\) \cite{wmap9}. 

While \(\partial^3 \eta/\partial M \partial q \partial z\) can be calculated using the pre-merger redshift \(z'\) we need the post-merger redshift \(z\) for the other terms in the integral in equation~\eqref{strain_integral}. This is found from the pre-merger redshift \(z'\) and the galaxy merger time \(T_{\text{gal-gal}}\)  by numerically solving for \(z_\mathrm{post-merger}\) via 
\begin{equation}
    T_{\text{gal-gal}} = \int_{t}^{t + T_{\text{gal-gal}}} dt = 
    \int_{z_\mathrm{post-merger}}^{z_\mathrm{pre-merger}}(1 + z) H(z) \, dz \, .
\end{equation}
With this, we are now fully equipped to calculate the characteristic strain spectrum \(h_c(f)\).

\begin{figure*}[tb]
  \centering
  \includegraphics[width=0.95\textwidth]{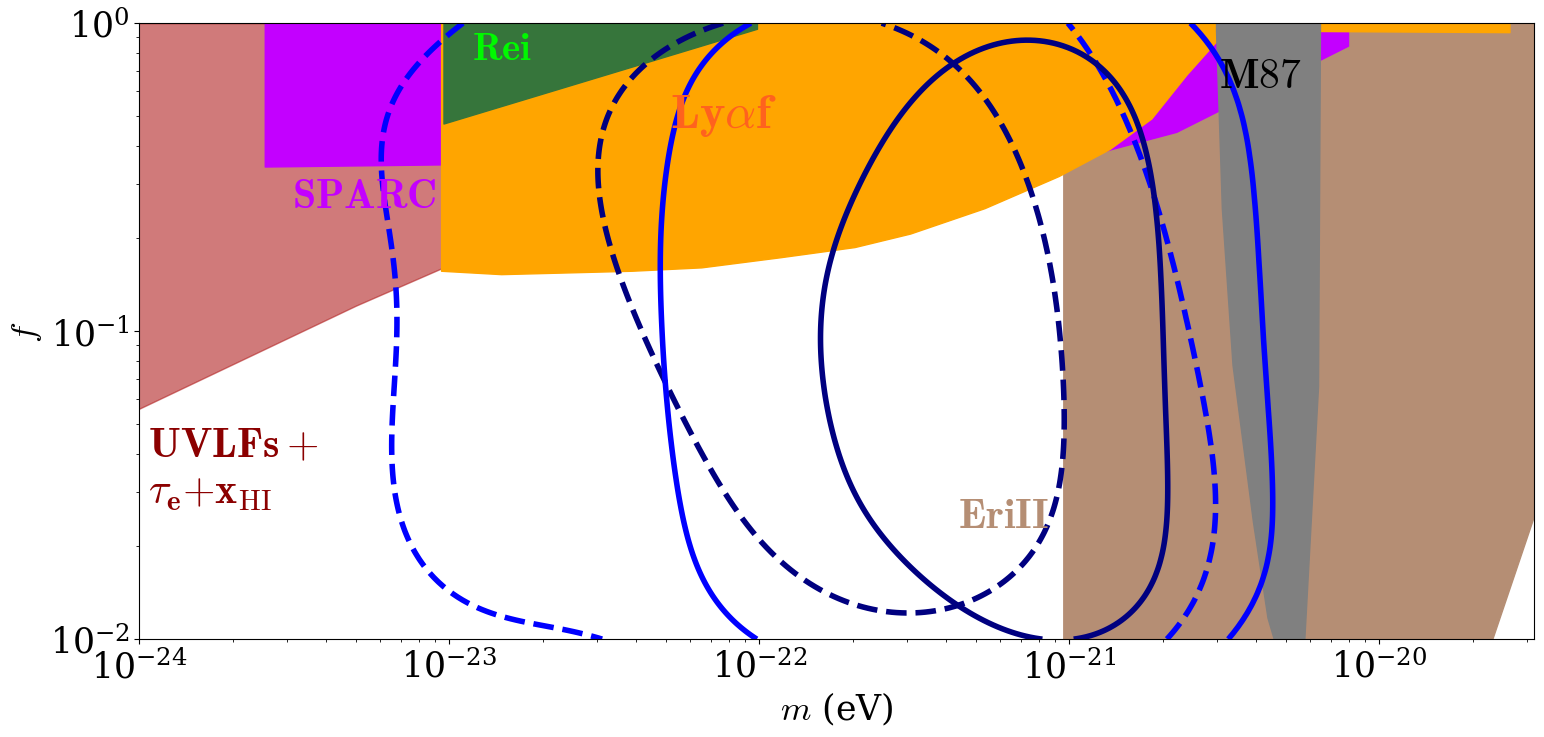}
  \caption{
The joint posterior for the ULDM particle mass and fraction is shown alongside existing constraints, as summarized in Ref.~\cite{Lazare:2024uvj}. 
Dark blue and blue lines show 68 and 95\% confidence intervals, respectively; the realistic and simple models are shown with solid and dashed lines, respectively.    }
  \label{fig:compare}
\end{figure*}

\section{ \label{sec:param_est} Parameter Estimation }

Our ULDM models, both simple and realistic, have six free parameters. The ULDM specific parameters are the particle mass \(m\), which we express as \(m / 10^{-22}\mathrm{eV} \equiv m_{22}\) and the ULDM fraction,  defined in Section \ref{subsec:dddt}.  The other parameters are contained within the galaxy stellar mass function (GSMF) \(\Psi(M_{\star}, z')\); \(\psi_0\) and \(m_{\psi, 0}\) from equation~\eqref{gsmf}, and 
the relation between $M_{\rm BH}$ and $M_{\rm bulge}$ 
(MMB), 
expressed in equation~\eqref{mbh_mbulge}; \(\mu\) and \(\epsilon_{\mu}\), following  Agazie2023. The priors are listed in Table \ref{table_param_values}.

We generate a library of gravitational strain spectra for SMBH binary mergers occurring within ULDM halos with different parameter combinations. We created 3000 unique combinations of the 6 model parameters following their prior distributions and ranges using Latin hypercube sampling.  We generated 2000 realizations of strain spectra for each of the 2000 parameter combinations, as detailed in Section V-C of Ref.~\cite{Tiruvaskar:2025lkq} and Section~3.2.2 of Agazie2023. 

We then train a Gaussian process (GP) on the library, allowing us to interpolate strain values at any intermediate point in the parameter space, speeding up MCMC generation. We train one GP for the median values of the strain and another for the standard deviation. The GP predictions and calculated strains differ by less than 2-\(\sigma\), confirming that the sample size is sufficient. 

The third and final step is MCMC creation. We used the NANOGrav 15-year dataset with Hellings-Downs correlated free spectrum modelled simultaneously with additional MP (monopole-correlated), DP (dipole-correlated) red noise, and CURN (common uncorrelated red noise). We only used the strain values from the five lowest frequency bins of this dataset, as these dominate the fit to PTA data. These strain values (or, more accurately, their distributions) are used to calculate likelihoods and thus posterior distributions from  MCMC chains. We use \texttt{holodeck} with modifications for our ULDM models for library generation, Gaussian process interpolation, and MCMC chain generation.

\section{\label{sec:results} Results and Analysis}

The posterior distributions for the ``simple'' and ``realistic'' models are presented in Figures~\ref{fig:posterior1} and \ref{fig:posterior2}. The 95\% region for  \(\psi_0\) matches well with the results of Ref.~\cite{alonso} who found a best fit of $\psi_0 \sim-2.5$. Our results for (\(\psi_0\), \(m_{\psi, 0}\), \(\mu\), \(\epsilon_{\mu}\)) are mainly prior driven and are in agreement with the results from those in Figure~10  of  Agazie2023, a phenomenological model with astrophysical priors and with Ref.~\cite{Tiruvaskar:2025lkq} for the self-interacting dark matter model. 

For the realistic model, the median  and 68\% credible interval  of the ULDM particle mass \(m\) is \(6.53 ^{+ 6.7} _{-4.6} \times 10^{-22}\) eV. For the simple model the allowed range is \(1.78 ^{+ 3.0} _{-1.3} \times 10^{-22}\) eV.   Figure~\ref{fig:compare} presents the 68\% and 95\% confidence parameter regions of our models, along with constraints on the ULDM parameter space \cite{Lazare:2024uvj}, and we see that these overlap with significant regions that are permitted by observations. For both models, the ULDM particle mass is well constrained, but the fraction is not, confirming the expectations of Section \ref{subsec:scaling}.

\begin{figure}
    \centering
    \includegraphics[width=\linewidth]{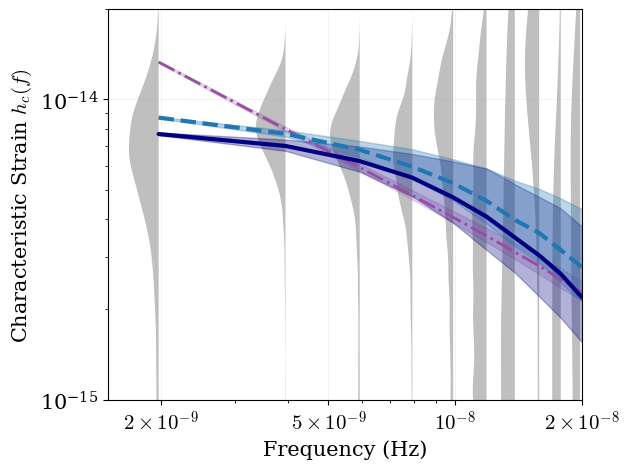}
    \caption{The dark blue line corresponds to the best-fit spectra for the realistic ULDM model; the dashed blue line represents the simple model. The purple line is the gravitational wave-only model from Agazie2023. Grey violins are the 15-year NANOGrav data. Shaded regions around the lines highlight the 95\% credible regions.}
    \label{fig:strain}
\end{figure}

Figure~\ref{fig:strain} shows the best-fit gravitational strain spectra.  The corresponding parameter values at the maximum log-posterior are given in Table~\ref{table_bestfit_values}.
\begin{table}[tb]
    \centering
    {
    \renewcommand{\arraystretch}{1.5}
    \begin{tabular}{|c|c|c|c|c|c|c|}
        \hline
        \textbf{Model} & $\boldsymbol{\psi_0}$ & $\boldsymbol{m_{\psi,0}}$ & $\boldsymbol{\mu}$ & $\boldsymbol{\epsilon_{\mu}}$ & $\boldsymbol{\log_{10}\!\left(\frac{m}{10^{-22}\,\mathrm{eV}}\right)}$ & $\boldsymbol{\log_{10} f}$ \\
        \hline
        Simple & $-2.33$ & $11.29$ & $8.61$ & $0.39$ & $0.25$ & $-0.68$ \\
        Realistic & $-2.66$ & $11.43$ & $8.62$ & $0.42$ & $0.87$ & $-0.52$ \\
        \hline
    \end{tabular}
    }
\caption{Maximum log-posterior values for the simple and realistic models.}
\label{table_bestfit_values}
\end{table}

\begin{figure*}[tb!]

\begin{subfigure}{.49\textwidth}
  \centering
  \includegraphics[width=.95\linewidth]{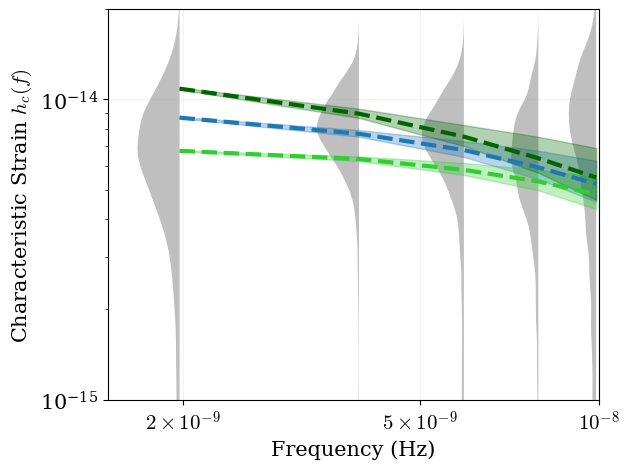}
  \label{fig:sfig2}
\end{subfigure}
\begin{subfigure}{.49\textwidth}
  \centering
  \includegraphics[width=.95\linewidth]{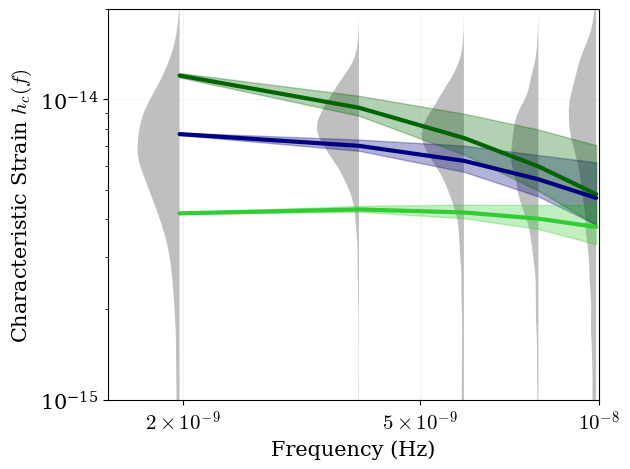}
  \label{fig:sfig1}
\end{subfigure}%
%

\caption{Strain spectra for a range of particle masses in both the simple (left- dashed lines) and realistic (right- solid lines) models. Blue lines are the best-fit spectra. Dark green lines correspond to the spectra with best-fit \(\mathrm{log}_{10}\left(\frac{m}{10^{-22}eV}\right)-0.25\) and the light green to best-fit \(\mathrm{log}_{10}\left(\frac{m}{10^{-22}eV}\right)+0.25\).}
\label{fig:varied_strain_spectra}
\end{figure*}

Spectra for different values of \(\mathrm{log}_{10}\left(\frac{m}{10^{-22}eV}\right)\) can be seen in Figures \ref{fig:varied_strain_spectra}. We find that in the realistic model, the strain is more sensitive to changes in the particle mass than in the simplified model, as expected from our scaling expectations.

We check that the black holes are inside the core radius as they contribute to the PTA signal, to confirm the assumptions used to compute ${dD}/{dt}$. Recalling that $f_{gw}$ is twice the orbital frequency, Kepler's third law gives 
\begin{equation}
    D=\left(\frac{GM_{BH}}{\pi^2f_{\mathrm{gw-source}}^2}\right)^{\frac{1}{3}}, 
    \label{eq:kepler} 
\end{equation}
where we have here neglected the piece of the soliton enclosed within the orbit.

\begin{figure}[tb]
    {\centering
    \includegraphics[width=\linewidth]{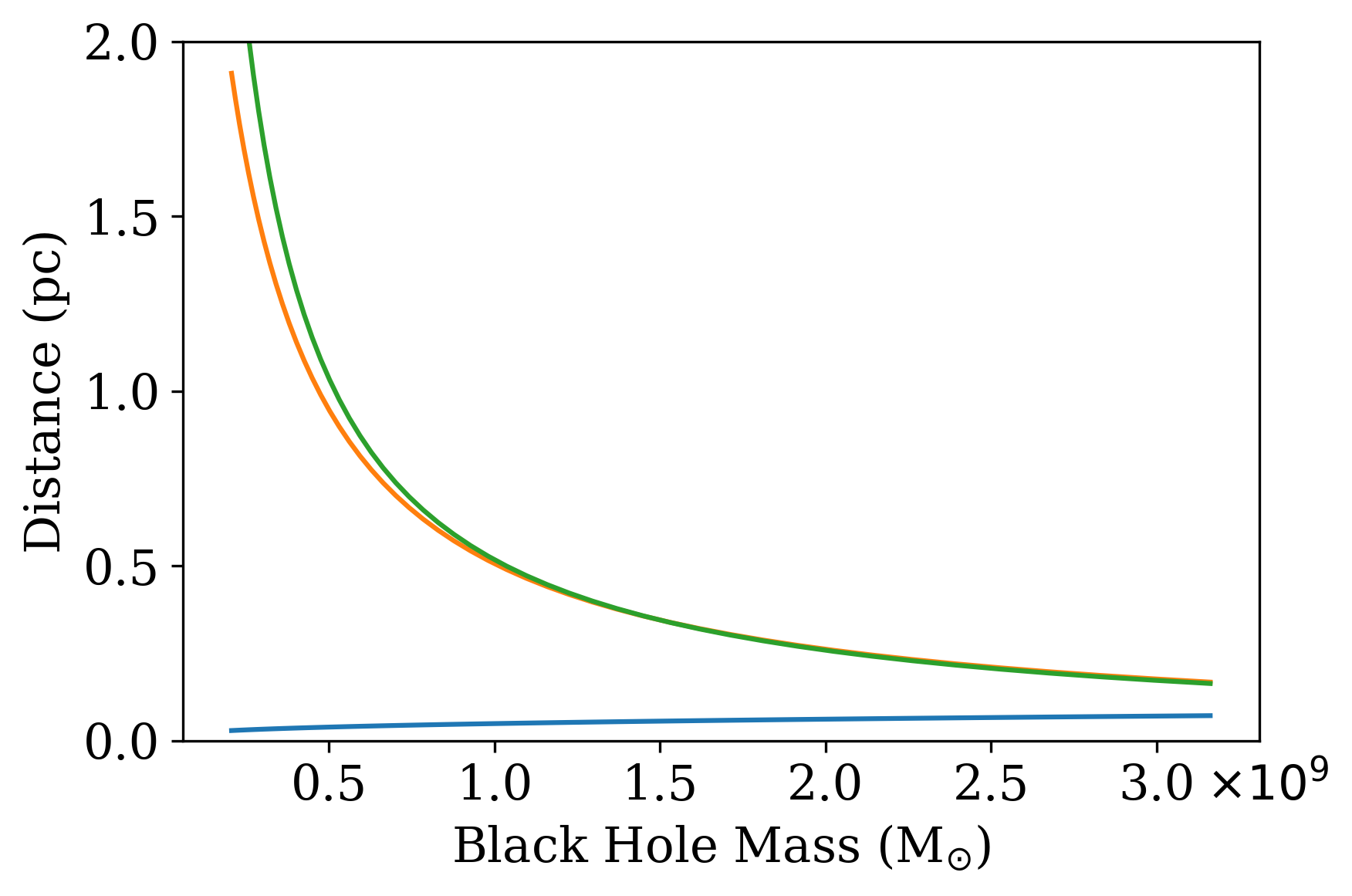}}
    \caption{Comparison of the soliton core radius at z=10 (orange) and z=0 (green) with the maximum separation of the binary where it emits in the PTA band (blue). This represents the maximum possible value that orbital radius can take for extreme binary mass ratios, which in practice do not produce large strain contributions.}
    \label{fig:r_c_comparison}
\end{figure}

Substituting $f_{\mathrm{gw-source}}=10^{-9}$Hz gives the maximum relevant separation at which the binary radiates at PTA frequencies. Figure \ref{fig:r_c_comparison} compares this to $r_c$ for  SMBH binaries that primarily contribute to our signal in the realistic model, and all of them are inside $r_c$. In the simplified model, there is no pinching, and $r_c$ is much larger than the orbital radius. 

It would take a full numerical simulation to confirm that solitons form in the halos which host the largest SMBH for a given ULDM fraction and particle mass, which is beyond the scope of this work. However, one consistency check is to compare the density of the naive soliton profile to that of the NFW profile of the cold dark matter component. We use the mass-concentration relationship in Ref~\cite{Child:2018skq} for the CDM component of the halo. If the soliton density is greater than the CDM density, it is more reasonable to assume soliton formation than otherwise. 

\begin{figure}[tb]
    \centering
    \includegraphics[width=\linewidth]{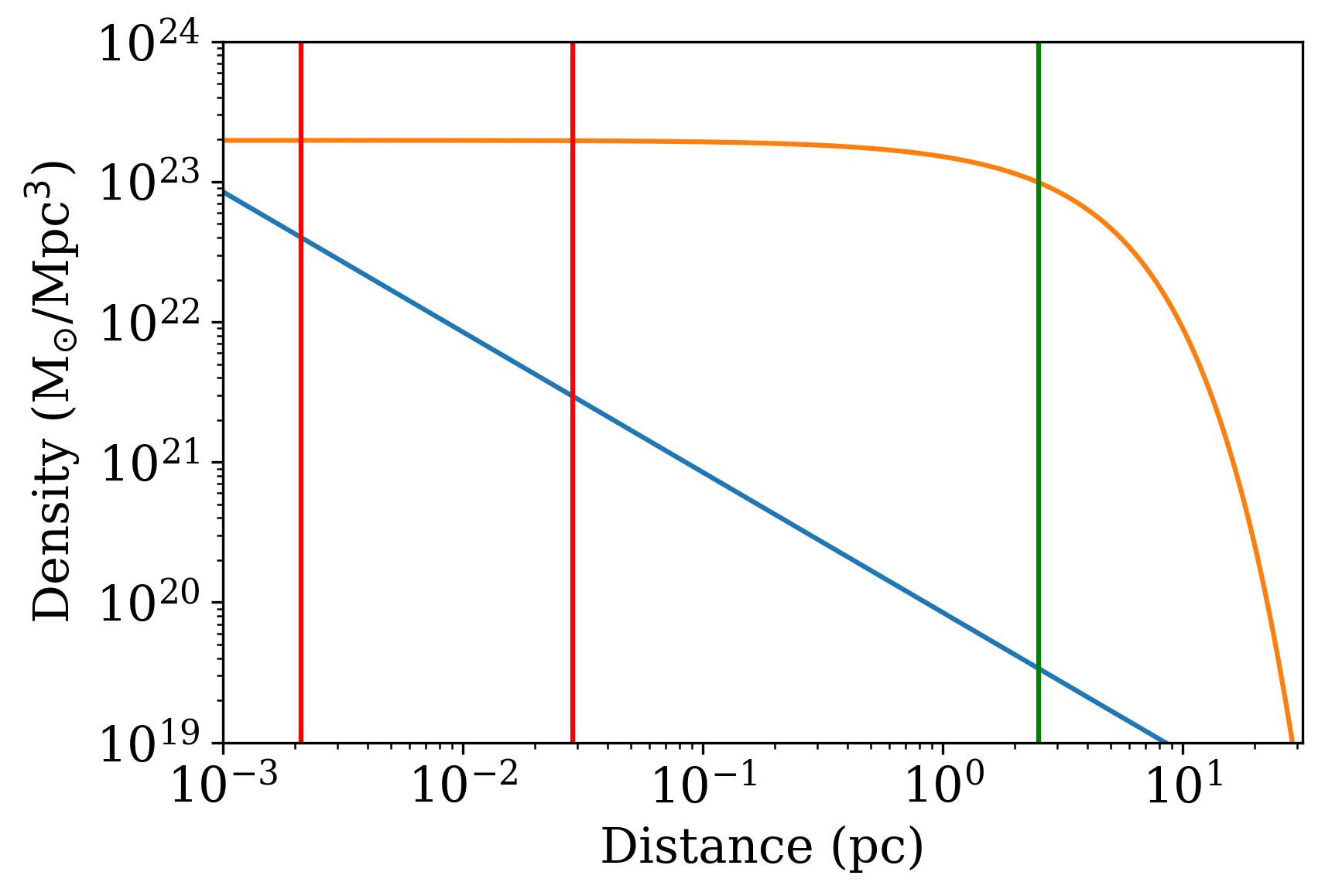}
    \caption{Comparisons of the soliton (orange) and NFW profile (blue) densities for a galaxy corresponding to $M_{\mathrm{BH}}=2.06\times10^{8}\text{M}_{\odot}$ at z=0. Red vertical lines correspond to separations associated with gravitational wave emission in the PTA band, while the green vertical line is the core radius of the soliton.}
    \label{fig:NFW_comparison}
\end{figure}

Figure \ref{fig:NFW_comparison} makes this comparison for our maximum likelihood parameters and a halo with $10^8 M_\odot$ black holes, the mass range that makes the largest contribution to the PTA signal. The soliton density exceeds the NFW density when pinching is accounted for, although it may not be satisfied in the simple model. Both the soliton and NFW profiles are affected by changing halo masses at higher redshifts, but the soliton mass generally increases with redshift directly in the core-halo relation, while the NFW concentration plateaus, so this is expected to hold at higher redshifts.

\section{\label{sec:discussion} Discussion and Conclusion}

We used the gravitational wave background detected in pulsar timing experiments to constrain the ULDM particle mass. To do this, we considered mergers of supermassive black holes in ULDM halos with solitonic cores,  developing a semi-analytic model of the dynamical friction experienced by an SMBH binary inside a soliton. 

We have confirmed that ULDM can contribute significantly to the orbital decay of SMBH binaries at the centers of the largest galaxies. The additional friction remains significant as binaries enter the pulsar timing band, resolving any ``final parsec problem" while suppressing the low-frequency component of the stochastic background,  improving the fit to current data.  A significant fraction of the ULDM parameter space is ruled out by a variety of astrophysical constraints \cite{Lazare:2024uvj}.  However, much of the region inside the joint posterior distribution for the ULDM parameters is consistent with all currently available constraints. Moreover, this shows that future pulsar timing measurements of the stochastic gravitational wave background will significantly constrain any ULDM contribution to the dark matter component of the universe. 

This work is based on an improved semi-analytic model for the dynamical friction on an SMBH binary in a ULDM soliton. This includes a heuristic rescaling of the soliton mass, accounting for both uncertainty in the core-halo relation and scenarios with a fractional ULDM dark matter contribution. Interestingly, scalings within the model imply that the dynamical friction from ULDM decreases slowly with the ULDM fraction. As a result, the contribution to binary decay remains significant even in scenarios with a relatively small quantity of ULDM. 

Because we work directly with a galaxy merger history we can identify the systems which make the largest contribution to the stochastic background; these are the halos in which the SMBH pair has roughly twice the mass of the soliton. This contrasts with previous work on ULDM-SMBH interactions by Aghaie \textit{et al.} \cite{Aghaie:2023lan}, which only considers systems where the black hole mass is at least five times larger than the soliton mass but our results are consistent with and complementary to theirs.  Moreover, constraints on astrophysical parameters from our analyses are in excellent agreement with previous results \cite{NANOGrav:2023hfp,Tiruvaskar:2025lkq,alonso}, which serves as a check on our model.

We developed two models of the dynamical friction - the ``realistic'' scenario accounts for the SMBH binary pinching the soliton profile, whereas the ``simple'' one does not. The best-fit spectra for both models match the data with a similar level of quality. In both cases, scaling relations in the model mean that drag drops as a small, fractional power of the ULDM fraction. As a result, we cannot put a clear lower bound on the ULDM fraction. That said, the model relies on a number of hypotheses, and the soliton mass is set via the ``standard'' core-halo relation.  Critically, this relation is an almost entirely unexplored relationship in models with a central SMBH, baryons and a mixture of ULDM and CDM. Each of these questions is an important avenue for future investigation. We have also assumed that there is a single ULDM component, which need not be the case, given that the arguments used to produce light scalar fields will typically be consistent with a spectrum with several such species \cite{Gosenca:2023yjc}. We have also ignored the soliton coalescence timescale in fractional ULDM models  \cite{Bar:2021kti}. In our case the halos are the products of mergers and $f$ itself is not directly equivalent to the ULDM fraction but this does suggest that there will be a lower bound on $f$ if additional drag is invoked to account for the PTA spectrum.

Constraints on the stochastic background are expected to improve substantially as higher-quality PTA data becomes available in the next few years. If the data continues to suggest an absence of power at low frequencies (relative to the simple gravitational-wave–only SMBH background prediction) PTA observations will identify a preferred region of the ULDM parameter space. Conversely, if this feature is not present in future datasets PTA observations will put stringent constraints on any possible ULDM contribution to the universe. 

In the relativistic regime, the possibility of the binary-ULDM system forming a so-called ``gravitational atom'' for certain ULDM mass parameters has been considered \cite{Baumann:2021fkf}, wherein excitation of the bound ULDM states produces an additional friction effect that may further impact the gravitational wave signal \cite{Guo:2025pea}. This additional effect may change our parameter constraints if added and is an interesting avenue for future work.  Conversely, Ref.~\cite{NANOGrav:2024nmo} considers the contribution of three-body scattering and eccentric orbits to the decay of SMBH binaries in the pulsar timing band and it will be critical to check that all  environment effects associated with conventional components of the universe are fully accounted for.

As noted above, the detailed drag models we use rely on several hypotheses.  However, both the simple and ``realistic'' models remove power from the stochastic background at low frequencies, giving confidence in our conclusion that PTA data is sensitive to the presence of ULDM in galaxies is robust. However, specific constraints will require a more detailed understanding of ULDM-SMBH interactions in recently merged galaxies.


\begin{acknowledgments}
We gratefully acknowledge support from the Marsden Fund Council grant MFP-UOA2131 from New Zealand Government funding, managed by the Royal Society Te Ap\={a}rangi, and  and we acknowledge
the use of New Zealand eScience Infrastructure (NeSI)
high-performance computing facilities. We thank Laura Burn, Emily Kendall, Renate Meyer, Pierre Mourier and Frank Wang for helpful discussions. We are grateful to Hovav Lazare for providing the code used to generate the other constraint contours in Fig.~\ref{fig:compare} and to the authors of \texttt{holodeck} for making the package available. We also acknowledge the University of Canterbury Research Cluster facilities for providing computational resources that significantly improved the efficiency of our computations (\href{https://doi.org/10.18124/CANTERBURYNZ-UCRCH}{DOI:10.18124/CANTERBURYNZ-UCRCH}, RRID:SCR\_027870). \end{acknowledgments}

\section*{Data availibility}
The code and the data used for producing the results of this paper are publicly avaliable \cite{tiruvaskar_2026_zenodo}.


%

\end{document}